\documentclass[10pt,a4paper]{article}

\usepackage{geometry}
\usepackage{multicol}
\usepackage{amsmath}
\usepackage{amsfonts}
\usepackage{mathtools}
\usepackage{bm}
\usepackage{fancyhdr}
\usepackage{hyperref}

\renewcommand{\sectionmark}[1]%
 {\markboth{\thesection:\ #1}{}}

\providecommand{\Ent}[1]{\lfloor #1 \rfloor}

\numberwithin{equation}{section}

\begin{document}

\thispagestyle{empty}

\title{\bfseries THE STRANGE HISTORY OF B FUNCTIONS \\
OR HOW THEORETICAL CHEMISTS \\ 
AND MATHEMATICIANS DO (NOT) INTERACT}

\author{Ernst Joachim Weniger \\
Institut f\"ur Physikalische und Theoretische Chemie \\
Universit\"at Regensburg, D-93040 Regensburg, Germany \\
Email: joachim.weniger@chemie.uni-regensburg.de}

\date{Submitted to International Journal of Quantum Chemistry:
  30 September 2008 \\
  Accepted for Publication: 11 November 2008}

\maketitle

\begin{abstract}
  \noindent
  $B$ functions are a class of relatively complicated exponentially
  decaying basis functions. Since the molecular multicenter integrals of
  the much simpler Slater-type functions are notoriously difficult, it is
  not at all obvious why $B$ functions should offer any advantages.
  However, $B$ functions have Fourier transforms of exceptional
  simplicity, which greatly simplifies many of their molecular
  multicenter integrals. This article discusses the historical
  development of $B$ functions from the perspective of the interaction
  between mathematics and theoretical chemistry, which traditionally has
  not been very good. Nevertheless, future progress in theoretical
  chemistry depends very much on a fertile interaction with neighboring
  disciplines.
\end{abstract}

\noindent {\bfseries Keywords:} Electronic structure theory,
exponentially decaying basis functions, $B$ functions, multicenter
integrals, interdisciplinary collaboration.


\tableofcontents

\newpage


\begin{multicols}{2}

\typeout{==> Section: Introduction}
\section{Introduction}
\label{Sec:Introduction}

Molecular electronic structure theory tries to explain the nature of
molecules and their properties by doing calculations on the basis of
quantum mechanics. This approach is obviously possible, but as we all
know, numerous obstacles have to be overcome. Already in 1929, Dirac had
stated \cite[p.\ 714]{Dirac/1929}:
\begin{quote}
  \sl The fundamental laws necessary for the mathematical treatment of a
  large part of physics and the whole of chemistry are thus completely
  known, and the difficulty lies only in the fact that application of
  these laws leads to equations that are too complex to be soluble. It
  therefore becomes desirable that approximate practical methods of
  applying quantum mechanics should be developed, which can lead to an
  explanation of the main features of complex atomic systems without too
  much computation. 
\end{quote}
It is noteworthy that Dirac only mentioned atomic calculations.
Apparently it was at that time more or less unthinkable to do meaningful
molecular calculations. But since the time of Dirac, there has been
enormous progress in particular in numerical and applied mathematics.
Moreover, we have witnessed a computer revolution which now gives us a
computing power that was beyond imagination in 1929. Accordingly, we now
have much better chances of accomplishing our aims. Chemical insight via
quantum mechanical calculations is no longer utopia.

Nevertheless, the central methodological problem addressed by Dirac --
the development of powerful approximation techniques -- has not really
changed. The reduction of chemical phenomena to \emph{manageable}
computational problems has been and still is far from trivial. In spite
of the theoretical and technological advances of recent years, the
computational complexity of sufficiently accurate quantum mechanical
calculations is still prohibitive, and molecular electronic structure
calculations are only feasible if we make some partly very drastic
approximations. 

Thus, progress depends quite crucially on our ability of reformulating
and translating chemical intuition to a mathematical formalism, which
does better than the currently available approximation procedures and
which ultimately helps to make more accurate calculations on larger
molecules possible.

Optimists might argue that future advances in computer hard- and software
will be enough to achieve this goal. However, I am not so optimistic. In
the long run, we will also need substantial progress in mathematics. But
this will to a large extend depend on a fruitful collaboration between
theoretical chemists interested in methodological questions and
mathematicians interested in scientific applications.

In theoretical chemistry, there is the widespread believe that the
parentage of atoms facilitates our attempts of understanding the
electronic structure of molecules. Thus, in molecular calculations on the
basis of the Hartree-Fock-Roothaan equations
\cite{Hartree/1928,Fock/1930,Roothaan/1951}, effective one-particle wave
functions (MOs) are usually approximated by a linear combination of
so-called \emph{basis functions} or \emph{atomic orbitals} (AOs) centered
at the $\Omega$ different nuclei of the molecule:
\begin{equation}
  \label{LCAO_MO_ansatz}
  \psi (\bm{r}) \; = \; \sum_{\alpha=1}^{\Omega} \, \sum_{n=1}^{N} \,
  C_{n}^{\alpha} \, \varphi_{n} (\bm{r}-\bm{R}_{\alpha}) \, .
\end{equation}

The fundamental entities in this LCAO-MO approach are the basis functions
$\{ \varphi_{n} (\bm{r}) \}_{n=1}^{N}$. Basis functions and their
derivatives up to a certain order must be square integrable with respect
to an integration over the whole $\mathbb{R}^{3}$, and they also have to
be complete in the corresponding Hilbert and Sobolev spaces, but
otherwise they are essentially arbitrary.

There are, however, two additional requirements of utmost practical
relevance which \emph{good} basis functions should satisfy:
\begin{enumerate}
\item Already short linear combinations of the type of
  (\ref{LCAO_MO_ansatz}) should provide sufficiently accurate
  approximations, i.e., the summation limit $N$ in (\ref{LCAO_MO_ansatz})
  should be as small as possible.
\item The molecular multicenter integrals, which occur inevitably in the
  LCAO-MO approach, should possess a manageable complexity.
\end{enumerate}

It is not too difficult to satisfy the first requirement by taking into
account what we know about the exact solutions of effective one-particle
atomic and molecular Schr\"{o}dinger equations. Kato \cite{Kato/1957} had
shown that the singularities of the potential of atomic and molecular
Hamiltonians produce discontinuities of the wave functions commonly
called Coulomb or correlation cusps. Moreover, exact solutions of these
Hamiltonians decay \emph{exponentially} at large distances (see for
example \cite{Agmon/1982,Agmon/1985} and references therein).

These properties of exact solutions clearly favor Slater-type functions
\cite{Slater/1930,Slater/1932}, which were originally introduced to
provide computationally convenient approximations to numerically
determined solutions of effective one-particle equations. In unnormalized
form, they can be expressed as follows:
\begin{equation}
  \label{Def_STF}
\chi_{N, L}^{M} (\beta, \mathbf{r}) \; = \;
(\beta r)^{N-L-1} \, \mathrm{e}^{- \beta r} \,
\mathcal{Y}_{L}^{M} (\beta \bm{r}) \, .
\end{equation}
Here, $\bm{r} \in \mathbb{R}^{3}$, $\mathcal{Y}_{L}^{M} (\beta \bm{r}) =
(\beta r)^{L} Y_{L}^{M} (\theta, \phi)$ is a regular solid harmonic and
$Y_{L}^{M} (\theta, \phi)$ is a (surface) spherical harmonic, $\beta > 0$
is a scaling parameter, $N$ is a generalized principal quantum number
which is in most cases, but not always a positive integer $\geq L+1$, and
$L$ and $M$ are the usual (orbital) angular momentum quantum numbers.

Slater-type functions were used with considerable success in atomic
calculation, and there is no reasonable doubt that their use in molecular
calculations would also be highly desirable. Unfortunately, Slater-type
functions are not able to satisfy the second requirement mentioned above:
The evaluation of their molecular multicenter integrals turned out to be
extremely difficult, and in spite of the heroic efforts of numerous
researchers, no completely satisfactory solution has been found yet.
Actually, the efficient and reliable evaluation of multicenter integrals
of Slater-type functions is among the oldest mathematical and
computational problems of molecular electronic structure theory. A review
of the older literature can be found in an article by Dalgarno
\cite{Dalgarno/1954}. Another useful review is the one by Harris and
Michels \cite{Harris/Michels/1967}.

Because of the problems with the efficient and reliable evaluation of the
notorious multicenter integrals of Slater-type functions, molecular
electronic structure calculations are now dominated by Gaussian basis
functions which have many disadvantages: They are unphysical because they
cannot reproduce a Coulomb cusp, and for large distances they also decay
much faster than exponentially. Accordingly, fairly large values of $N$
in (\ref{LCAO_MO_ansatz}) are needed to produce sufficiently accurate
results. In fact, Gaussian basis functions have only one, albeit decisive
advantage: Their molecular multicenter integrals can be computed
comparatively easily.

In spite of the clear dominance of Gaussian functions, the search for
alternative basis functions, that are more physical than Gaussians and
whose multicenter integrals can be computed more easily than the
notoriously difficult multicenter integrals of Slater-type functions, has
been continued by numerous researchers. In this article, I want to
discuss the history and most interesting features of a class of
exponentially decaying basis functions, the so-called $B$ functions, that
were -- based on previous work by Shavitt \cite[Eq.\ (55) on p.\
15]{Shavitt/1963} -- defined by Filter and Steinborn \cite[Eq.\
(2.14)]{Filter/Steinborn/1978b} as follows:
\begin{equation}
  \label{Def:B_Fun}
  B_{n,\ell}^{m} (\beta, \hm{r}) \; = \; \frac{\hat{k}_{n-1/2} (\beta r) \, 
    \mathcal{Y}_{\ell}^{m} (\beta \bm{r})}{2^{n+\ell} (n+\ell)!} \, . 
\end{equation}
Here, $\beta > 0$, $n \in \mathbb{Z}$, and $\hat{k}_{n-1/2}$ is a reduced
Bessel function. If $K_{\nu} (z)$ is a modified Bessel function of the
second kind \cite[p.\ 66]{Magnus/Oberhettinger/Soni/1966}, the reduced
Bessel function with in general complex $\nu$ and $z$ is defined as
follows \cite[Eqs.\ (3.1) and (3.2)]{Steinborn/Filter/1975c}
\begin{equation}
   \label{Def:RBF}
\hat{k}_{\nu} (z) \; = \; (2/\pi)^{1/2} \, z^{\nu} \, K_{\nu} (z) \, .
\end{equation}
If the order $\nu$ is half-integral, $\nu = n + 1/2$ with $n \in
\mathbb{N}_0$, a substantial simplification takes place: Then, the
reduced Bessel function is an exponential multiplied by a terminating
confluent hypergeometric series ${}_1 F_1$ (see for example \cite[Eq.\
(3.7)]{Weniger/Steinborn/1983b}):
\begin{align}
  \label{RBF_HalfInt}
  & \hat{k}_{n+1/2} (z) 
  \notag 
  \\ 
  & \qquad \; = \; 2^n \, (1/2)_n \,
  \mathrm{e}^{-z} \, {}_1 F_1 (-n; -2n; 2z) \, .
\end{align}
Obviously, $B$ functions are fairly complicated mathematical objects.
Moreover, (\ref{Def:B_Fun}) and (\ref{RBF_HalfInt}) imply that $B$
functions can be expressed as a linear combination of Slater-type
functions defined by (\ref{Def_STF}). Hence, it is not at all clear why
the use of the comparatively complicated $B$ functions should offer any
advantages over Slater-type functions which possess an exceptionally
simple explicit expression. 

If we form finite linear combinations of Slater-type functions and do
some mathematical manipulations, then the complexity of the resulting
expression normally increases, depending on the number of Slater-type
functions occurring in the linear combination.  In fortunate cases,
however, it may happen that most terms of the resulting expression cancel
exactly. Thus, a significant reduction of complexity by forming linear
combinations is also possible.

The topic of this article are $B$ functions and their role in the theory
of multicenter integrals. However, research on multicenter integrals is
essentially mathematical in nature. Quantum mechanics only determines
which integrals we evaluate, but the techniques employed for their
evaluations are entirely mathematical. Physical insight is of secondary
importance. What really matters are mathematical skills.

Thus, a discussion of $B$ functions immediately raises another, very
profound question about the role of mathematics in molecular electronic
structure theory, and how well and how rapidly new advances in
mathematics are incorporated into electronic structure theory.

Essentially this boils down to the question whether and how well
theoretical chemists interested in methodological questions and
mathematicians are interacting and -- in the best of all worlds -- even
collaborating. As I had argued in \cite[Section 1]{Weniger/2005}, science
is now highly fragmented and consists of almost completely disjoint
subfields. As a consequences of this probably unavoidable specialization,
communication between researchers belonging to different subfields has
deteriorated quite a bit and no longer functions as well as it should.

Molecular electronic structure theory is a highly interdisciplinary
research topic whose progress depends crucially on a functioning
communication with and input from neighboring disciplines. Ultimately,
computational chemistry has become feasible because of the spectacular
advances in computer hard- and software. However, mathematics is the
basis of all computing and not all problems can be solved by using more
powerful computers. One should also not forget that Pople, whose suite of
Gaussian programs has contributed greatly to the computerization of
chemistry, had done his undergraduate studies in mathematics, and later
he had been reader in mathematics at the University of Cambridge.

It is my central hypothesis that it should be in the self-interest of
theoretical chemists to improve communication with mathematicians.
Traditionally, this communication has been fairly bad. In my opinion, the
history of $B$ functions provides ample evidence supporting this claim.

\typeout{==> Section: Bessel Polynomials}
\section{Bessel Polynomials}
\label{Sec:BesselPolynomials}

Polynomials associated with the names of Legendre, Gegenbauer, Jacobi,
Hermite, and Laguerre certainly belong to the pillars of classical
analysis, and they are also of considerable importance in quantum
theory and other scientific disciplines. 

There are several other, less important classes of polynomials with well
established properties. Many of these polynomials are completely ignored
by theoretical physicists or chemists, because they have no obvious
physical relevance. Such a pragmatic attitude is certainly justified.
Nevertheless, occasional surprises cannot be ruled out.

For example, in the later stages of my PhD thesis \cite{Weniger/1982} I
became aware of a book by Grosswald \cite{Grosswald/1978} from which I
learned that the polynomial part in (\ref{RBF_HalfInt}) and thus also
reduced Bessel functions had already been treated in the mathematical
literature. For the polynomial part of reduced Bessel functions, the
notation
\begin{equation}
  \label{Def:BesPol}
  \theta_n (z) \; = \; \mathrm{e}^z \, \hat{k}_{n+1/2} (z)
\end{equation}
is used in the mathematical literature \cite[Eq.\ (1) on p.\
34]{Grosswald/1978}. Together with some other, closely related
polynomials, the $\theta_n (z)$ are called \emph{Bessel polynomials}
\cite{Grosswald/1978}. According to Grosswald \cite[Section
14]{Grosswald/1978}, they have been applied in such diverse and seemingly
unrelated fields as number theory, statistics, and the analysis of
complex electrical networks.

The Bessel polynomials $\theta_n (z)$ occur also in a completely
different mathematical context which later became very important for me.
In the book by Baker and Graves-Morris \cite[p.\
8]{Baker/Graves-Morris/1996} on Pad\'{e} approximants, it is remarked
that Pad\'{e} had shown in his seminal thesis \cite{Pade/1892} that the
Pad\'{e} approximant $[n/m]$ with $n, m \in \mathbb{N}_{0}$ to the
exponential function $\exp (z)$ can be expressed as the ratio of two
terminating confluent hypergeometric series \cite[Eq.\
(2.12)]{Baker/Graves-Morris/1996}:
\begin{equation} 
  [n/m] \; = \; \frac
   {{}_1 F_1 (- n; - n - m; z)}
   {{}_1 F_1 (- m; - n - m; - z)} \, .
\end{equation}
Accordingly, the diagonal Pad\'{e} approximant with $n = m$ to the
exponential function can be expressed as the ratio of two Bessel
polynomials:
\begin{equation}
[n/n] \; = \; \frac{\theta_n (z/2)}{\theta_n (-z/2)} \, ,
\qquad n \in \mathbb{N}_0 \, .
\end{equation}

\typeout{==> Section: Shavitt's Gauss Integral Representation}
\section{Shavitt's Gauss Integral Representation}
\label{Sec:ShavittsGaussIntegralRepresentation}

We can safely assume that Bessel polynomials were completely unknown
among theoretical chemists. Thus, the usefulness of reduced Bessel
functions had to be discovered differently.

It seems that Shavitt was the first theoretical chemist who noticed that
reduced Bessel functions have in spite of their undeniable complexity
some very useful features. In tables of Laplace transforms, one can find
the following relationship (see for example \cite[Eq.\ (13.42) on p.\
338]{Oberhettinger/Badii/1973}):
\begin{equation}
  \label{LapIntRep_RBF}
  s^{\nu/2} \, K_{\nu} \bigl( a s^{1/2} \bigr) \; = \; a^{\nu} \,
  \int_{0}^{\infty} \, \frac{\mathrm{e}^{-st-a^2/(4t)}}{(2t)^{\nu+1}} \,
  \mathrm{d} t \, .
\end{equation}
Setting $s = z^2$ and $a = 1$ yields the following remarkably simple
Gaussian integral representation of the reduced Bessel function
\cite[Eq.\ (55) on p.\ 15]{Shavitt/1963}:
\begin{equation}
  \label{GaussIntRep_RBF}
\hat{k}_{\nu} (z) \; = \; (2/\pi)^{1/2} \,
\int_{0}^{\infty} \, \frac{\mathrm{e}^{-z^2 t-1/(4t)}}{(2t)^{\nu+1}} \,
\mathrm{d} t \, .
\end{equation}
Shavitt defined the reduced Bessel function according to $k_{\lambda} (z)
= z^{\lambda} K_{\lambda} (z)$. Unfortunately, this notation is
misleading because $k_{\lambda} (z)$ can easily be confounded with the
\emph{spherical} modified Bessel function $k_{n} (z) = [\pi/(2z)]^{1/2}
K_{n+1/2} (z)$ with $n \in \mathbb{N}_{0}$. Therefore, Steinborn and
Filter \cite[Eqs.\ (3.1) and (3.2)]{Steinborn/Filter/1975c} defined
reduced Bessel functions via (\ref{Def:RBF}). The introduction of the
prefactor $(2/\pi)^{1/2}$ in (\ref{Def:RBF}) has the additional advantage
that it simplifies most formulas.

A Gauss integral representation for Slater-type functions can also be
constructed. The starting point would be the following following Laplace
transform (see for example \cite[Eq.\ (5.94) on p.\
259]{Oberhettinger/Badii/1973}):
\begin{align}
  \label{LapIntRep_STF}
  & s^{\nu} \, \exp \bigl( - a s^{1/2} \bigr) \; = \; [2/\pi]^{1/2} \,
  \notag
  \\
  & \quad \times \, 
  \int_{0}^{\infty} \, \frac{\mathrm{e}^{-st-a^2/(8t)}}{(2t)^{\nu+1}} \,
  D_{2\nu+1} \bigl( a/[2t]^{1/2} \bigr) \, \mathrm{d} t \, .  
\end{align}
Here, $D_{2\nu+1} \bigl( a/[2t]^{1/2} \bigr)$ is a parabolic cylinder
function \cite[p.\ 324]{Magnus/Oberhettinger/Soni/1966} which is a
special case of the Whittaker function $W_{\kappa, \mu} (z)$ of the
second kind \cite[p.\ 296]{Magnus/Oberhettinger/Soni/1966}. Accordingly,
the Laplace transform (\ref{LapIntRep_STF}) is much more complicated than
(\ref{LapIntRep_RBF}). Obviously, this conclusion applies also to the
Gauss transform for Slater-type functions that can be derived from
(\ref{LapIntRep_STF}).

The remarkably compact Gauss integral representation
(\ref{GaussIntRep_RBF}) inspired Shavitt \cite[p.\ 16]{Shavitt/1963} to
propose the use of reduced Bessel functions as alternative exponentially
decaying basis functions in electronic structure calculations. This was
done in articles by Bishop and Somorjai \cite{Bishop/Somorjai/1970},
Somorjai and Yue \cite{Somorjai/Yue/1970}, and Yue and Somorjai
\cite{Yue/Somorjai/1970}.

The Gaussian function in (\ref{GaussIntRep_RBF}) makes it possible to
reduce multicenter integrals involving reduced Bessel functions to the
corresponding integrals of Gaussians, which can be evaluated
comparatively easily, multiplied by integrals over \emph{nonphysical}
variables. Unfortunately, the remaining integrations can normally only be
done numerically. Therefore, this approach is apparently not efficient
enough to provide a viable approach for the evaluation of multicenter
integrals of reduced Bessel functions.

\typeout{==> Section: The Gegenbauer Addition Theorem of Reduced Bessel
  Functions}
\section{The Gegenbauer Addition Theorem of Reduced Bessel Functions}
\label{Sec:GegenbauerAdditionTheoremRBF}

Multicenter integrals are difficult to evaluate because the integration
variables are not separated. Principal tools, which can accomplish such a
separation of variables, are so-called \emph{addition theorems}. These
are expansions of a given function $f (\bm{r} \pm \bm{r}')$ with $\bm{r},
\bm{r}' \in \mathbb{R}^3$ in terms of other functions that only depend on
either $\bm{r}$ or $\bm{r}'$. The best known example of such an addition
theorem is the Laplace expansion of the Coulomb potential in terms of
spherical harmonics, which is nothing but the well known generating
function $\bigl[1-2xz+z^{2}\bigr]^{-1/2}$ of the Legendre polynomials
\cite[p.\ 232]{Magnus/Oberhettinger/Soni/1966} is disguise.

The modified Bessel function $w^{-\nu} K_{\nu} (\gamma w)$ with $ w =
\bigl[ \rho^{2} + r^{2} - 2 r \rho \cos \theta \bigr]^{1/2}$, $0 < \rho <
r$, and $\nu \in \mathbb{C} \setminus \mathbb{N}_{0}$ possesses the
following Gegenbauer-type addition theorem \cite[pp.\ 106 -
107]{Magnus/Oberhettinger/Soni/1966}:
\begin{align}
  \label{BesK_GegenbauerAddThm}
  & w^{-\nu} K_{\nu} (\gamma w) \; = \; 2^{\nu} \, \gamma^{-\nu} \,
  \Gamma (\nu) \, (r \rho)^{-\nu}
   \notag \\
   & \quad \times \,
   \sum_{n=0}^{\infty} \, C_{n}^{\nu} (\cos \theta ) \,
   I_{\nu+n} (\gamma \rho) \, K_{\nu+n} (\gamma r) \, .
\end{align}
Here, $C_{n}^{\nu} (\cos \theta )$ is a Gegenbauer polynomial \cite[p.\
218]{Magnus/Oberhettinger/Soni/1966}, and $I_{\nu+n} (\gamma \rho)$ is a
modified Bessel function of the first kind \cite[p.\
66]{Magnus/Oberhettinger/Soni/1966}.

Comparison with (\ref{Def:RBF}) shows that the addition theorem
(\ref{BesK_GegenbauerAddThm}) is essentially a Gegenbauer-type addition
theorem of the reduced Bessel function $\hat{k}_{-\nu} (\gamma w)$. A
very important special case occurs for $\nu=1/2$ because then the
Gegenbauer polynomial becomes a Legendre polynomial according to
$C_{n}^{1/2} (x) = P_{n} (x)$ \cite[p.\
219]{Magnus/Oberhettinger/Soni/1966}, yielding the Legendre-type addition
theorem of the Yukawa potential \cite[p.\
218]{Magnus/Oberhettinger/Soni/1966}:
\begin{align}
  \label{Yukawa_LegAddThm}
  \frac{\mathrm{e}^{{- \gamma w}}}{w} & \; = \;(r \rho)^{-1/2} \,
  \sum_{n=0}^{\infty} \, (2n+1) \, P_{n} (\cos \theta ) 
  \notag
  \\
  & \quad \qquad \times \, 
  I_{n+1/2} (\gamma \rho) \, K_{n+1/2} (\gamma r) \, .
\end{align}
With the help of the so-called addition theorem of the Legendre
polynomials \cite[p.\ 303]{Biedenharn/Louck/1981a}
\begin{align}
  \label{AddLegPol}
  & P_{\ell} \bigl( \cos \theta \bigr) \; = \; \frac{4\pi}{2 \ell + 1}
  \notag 
  \\
  & \qquad \times \, \sum_{m=-\ell}^{\ell} \,
  \bigl[ Y_{\ell}^{m} (\bm{r}/r) \bigr]^{*} \, 
  Y_{\ell}^{m} (\bm{r}'/r') \, ,
\end{align}
where $\cos \theta = \bm{r} \bm{r}'/(r r')$, the Legendre-type addition
theorem (\ref{Yukawa_LegAddThm}) can be converted to an expansion in
terms of spherical harmonics.

This addition theorem for the Yukawa potential was the starting point for
the so-called zeta function method of Barnett and Coulson
\cite{Barnett/Coulson/1951}. They tried to construct an addition theorem
for Slater-type functions by applying suitable \emph{generating
  differential operators} to this addition theorem. Unfortunately, this
idea did not lead to a complete success in the case of Slater-type
functions. Due to technical problems it was not possible to perform all
differentiations in closed form. Consequently, the coefficients of the
zeta function expansion could only be computed recursively, but not in
explicit form \cite{Barnett/Coulson/1951,Barnett/1963}.

Later, numerous other researchers worked on addition theorems for
Slater-type functions (see for example \cite[Refs.\ 33 - 50, 52, 65, 84,
85, 90, 91]{Weniger/2002}). But all these addition theorems turned out to
be fairly complicated. I suspect that there is still plenty of room for
improvements.

The technical problems with the zeta function method of Barnett and
Coulson \cite{Barnett/Coulson/1951} in special and with addition theorems
for Slater-type functions in general inspired Steinborn and Filter to
derive an addition theorem for reduced Bessel functions directly from the
Gegenbauer expansion (\ref{BesK_GegenbauerAddThm}). This was much easier
than the derivation of analogous addition theorems for scalar Slater-type
functions.

Let us assume that a Gegenbauer-type addition of the type of
(\ref{BesK_GegenbauerAddThm}) for some function is known. If the
Gegenbauer polynomials in this expansion can be replaced by a finite sum
of Legendre polynomials, a rearrangement of the original expansion yields
an expansion in terms of Legendre polynomials. Then, we only need
(\ref{AddLegPol}) to obtain an expansion in terms of spherical harmonics.

The practical realization of this obvious idea was apparently not so
easy. As discussed by Steinborn and Filter \cite[pp.\ 269 -
270]{Steinborn/Filter/1975c}, many authors had quite a few problems with
the determination of explicit expressions for the coefficients of the
expansion of Gegenbauer polynomials in terms of Legendre polynomials.
Also Steinborn and Filter constructed very messy expressions for these
coefficients which are restricted to certain superscripts of the
Gegenbauer polynomial \cite[Section 3]{Steinborn/Filter/1975b}.

However, already at that time a much more convenient expression for these
expansion coefficients been available in the mathematical literature. In
Exercise 4 on p.\ 284 of Rainville's book \cite{Rainville/1971}, one
finds the following expansion of Gegenbauer polynomials in terms of
Legendre polynomials, but no explicit proof (compare also \cite[Eq.\
(5.2)]{Weniger/Steinborn/1989b}):
\begin{align}
  \label{Gegenbauer_2_Legendre}
  & C_{m}^{\mu} (x) \; = \; \sum_{s=0}^{\Ent{m/2}} \, \frac{(\mu)_{m-s}
    \, (\mu-1/2)_{s}}{(3/2)_{m-s} \, s!}  \notag
  \\
  & \qquad \times \, (2m - 4 s + 1) \, P_{m-2s} (x) \, .
\end{align}
Here, $\Ent{m/2}$ denotes the integral part of $m/2$.

Expansion (\ref{Gegenbauer_2_Legendre}) can be proved via the explicit
expression \cite[Eq.\ 7.313.7 on p.\ 836]{Gradshteyn/Rhyzhik/1994} for
the integral $\int_{-1}^{1} (1-x)^{\alpha} (1-z)^{\nu-1/2} C_{m}^{\mu}
(x) C_{n}^{\nu} (x) \mathrm{d} x$. One only has to set $\nu=1/2$ and
perform the limit $\alpha \to 0$, which requires, however, some algebraic
trickery.

In this way, the following addition theorem for a reduced Bessel function
with half-integral order can be derived \cite[Eq.\
(5.5)]{Weniger/Steinborn/1989b}):
\begin{align}
  \label{RBF_AddThm_HalfInt}
& \hat{k}_{n-1/2} \bigl(\beta \vert \bm{r}_{<} \pm \bm{r}_{>} \vert \bigr)
\notag 
\\ 
& \quad = \; \frac{(-1)^{n} 8\pi}{(2n-1)!!} \, 
(\beta r_<)^{n-1/2} (\beta r_>)^{n-1/2}
\notag 
\\
& \quad \times \, \sum_{\ell=0}^{\infty} \,  
\sum^{\ell}_{m=-\ell} \, (\mp 1)^{\ell} \,
\bigl[ Y^{m^{\star}}_{\ell}(\bm{r}_{<}/r_{<}) \bigr]^{*} \, 
Y^m_{\ell} (\bm{r}_{>}/r_{>}) 
\notag 
\\
& \quad \times \, 
\sum^{n}_{\nu=0} \, \frac{(-n)_{\nu} (1/2-n)_{\ell+\nu}}
{{\nu}! (3/2)_{\ell+\nu}} \, (\ell+2\nu-n+1/2) 
\notag
\\
& \qquad \times \,
I_{\ell+2\nu-n+1/2}(\beta r_{<})
K_{\ell+2\nu-n+1/2}(\beta r_{>}) \, .
\end{align}
Here, we have $\vert \bm{r}_{<} \vert < \vert \bm{r}_{>} \vert$. The
original version of this addition theorem as derived by Steinborn and
Filter \cite[Eq.\ (3.4)]{Steinborn/Filter/1975c} still contains
unspecified expansion coefficients of Gegenbauer polynomials in terms of
Legendre polynomials.

The addition theorem (\ref{RBF_AddThm_HalfInt}) was quite consequential
for my later scientific interests. In my diploma thesis
\cite{Weniger/1977}, which was published in condensed form in
\cite{Steinborn/Weniger/1977}, I used this addition theorem for the
evaluation of simple multicenter integrals of reduced Bessel functions.

\typeout{==> Section: Convolution Theorems of B Functions}
\section{Convolution Theorems of B Functions}
\label{Sec:ConvolutionTheorems_B_Fun}

The original topic of Filter's PhD thesis \cite{Filter/1978} was the
construction of addition theorems and their use for the evaluation of
multicenter integrals. During these investigations, it became obvious
that the use of addition theorems alone would not suffice to permit an
efficient and reliable evaluation of the complicated multicenter
integrals of exponentially decaying functions. Thus, alternative
approaches were necessary.

Starting from the well known addition theorem of the so-called modified
Helmholtz harmonics, Filter succeeded in his PhD thesis \cite[Sections 7
- 9]{Filter/1978} to derive with the help of some very skillful
mathematical manipulations remarkably compact expressions for
convolution, nuclear attraction, and Coulomb integrals of $B$ functions
which were later published in
\cite{Filter/Steinborn/1978b,Filter/Steinborn/1978a}.

Probably the most spectacular result is the extremely compact expression
for the overlap integral of two $B$ functions with equal scaling
parameters $\beta > 0$, which is simply a finite linear combination of
$B$ functions \cite[Eq.\ (4.3)]{Filter/Steinborn/1978b}:
\begin{align}
  \label{ConvInt_Bnlm_ESP}
  & \int \, 
  \bigl[ B_{n_1,\ell_1}^{m_1} (\beta, \bm{r}) \bigr]^{*} \, 
  B_{n_2, \ell_2}^{m_2} (\beta, \bm{r}-\bm{R}) \, \mathrm{d}^{3} \bm{r}
  \notag \\
  & \quad \; = \; \frac{4\pi}{\beta^3} \,
  \sum_{\ell=\ell_{\mathrm{min}}}^{\ell_{\mathrm{max}}} \! {}^{(2)} \,
  \langle \ell_2 m_2 \vert \ell_1 m_1 \vert \ell m_2-m_1 \rangle
  \notag \\
  & \qquad \times \, \sum_{t=0}^{\Delta \ell} \, (-1)^t \,
  {\binom{\Delta \ell} {t}}
  \notag \\
  & \quad \qquad \times \, B_{n_1+n_2+\ell_1+\ell_2-\ell-t+1,
    \ell}^{m_2-m_1} (\beta, \bm{R}) \, .
\end{align} 
Here, $\langle \ell_2 m_2 \vert \ell_1 m_1 \vert \ell m_2-m_1 \rangle$ is
a so-called Gaunt coefficient, which is the integral of three spherical
harmonics over the surface of the unit sphere in $\mathbb{R}^{3}$. A
compact review of the properties of Gaunt coefficients plus additional
references can be found in \cite[Appendix C]{Weniger/2005}. Moreover, it
follows from the selection rules satisfied by the Gaunt coefficient (see
for example \cite[Eq.\ (3.1)]{Weniger/Steinborn/1982}) that the summation
limit $\Delta \ell = (\ell_{1}+\ell_{2}-\ell)/2$ is a non-negative
integer.

For overlap integrals with different scaling parameters $\alpha \neq
\beta > 0$, compact finite expressions involving Jacobi polynomials could
be derived \cite[Eq.\ (4.6)]{Filter/Steinborn/1978b}. In addition, there
are infinite series representations expressing an overlap integral with
different scaling parameters as an infinite series of overlap integrals
with equal scaling parameters \cite[Eq.\ (4.9)]{Filter/Steinborn/1978b}.  

Explicit expressions of similar complexity could also be derived for
nuclear attraction integrals \cite[Eqs.\ (6.4) and
(6.5)]{Filter/Steinborn/1978b} and in particular also for Coulomb
integrals, which are six-dimensional integrals and which describe the
electrostatic interaction of one-center charge densities represented by
$B$ functions \cite[Eqs.\ (7.5), (7.7), (7.10), (7.13), (7.15), and
(7.16)]{Filter/Steinborn/1978b}

Since Slater-type functions $\chi_{N, L}^{M}$ with integral principal
quantum numbers $N \in \mathbb{N}$ can be expressed by finite linear
combinations of $B$ functions \cite[Eq.\ (6.3)]{Filter/Steinborn/1978a},
the remarkably compact formulas for overlap, nuclear attraction, and
Coulomb integrals of $B$ functions can be used to write down analogous
expressions for the corresponding integrals of Slater-type functions in a
straightforward way.

The extremely compact expression (\ref{ConvInt_Bnlm_ESP}) was the basis
for the construction of certain one-range addition theorems by Filter and
Steinborn \cite{Filter/Steinborn/1980}. These addition theorems are
expansions in terms of a complete and orthonormal function set based on
the generalized Laguerre polynomials. They converge in the mean with
respect to the norm of the Hilbert space $L^{2} (\mathbb{R}^{3})$ of
square integrable functions.

For the evaluation of the expansion coefficients of these addition
theorems, which are essentially overlap integrals, Filter and Steinborn
only had to combine (\ref{ConvInt_Bnlm_ESP}) with the following expansion
of generalized Laguerre polynomials in terms of reduced Bessel functions
\cite[Eq.\ (3.3-35)]{Weniger/1982} (see also \cite[Eq.\ (3.17) and Ref.\
\lbrack 23\rbrack\ on p.\ 2736]{Filter/Steinborn/1980}):
\begin{align}
  \label{GLag_FinSum_RBF}
  & \mathrm{e}^{-z} \, L_{n}^{(\alpha)} (2z) \; = \; (2n+\alpha+1)
  \notag \\
  & \quad \times \, 
  \sum_{\nu=0}^{n} \, \frac{(-2)^{\nu} \Gamma (n+\alpha+\nu+1)}
  {\nu! (n-\nu)! \Gamma (\alpha+2\nu+2)} \, \hat{k}_{\nu+1/2} (z) \, .
\end{align} 

\typeout{==> Section: The Fourier Transform of B Functions}
\section{The Fourier Transform of B Functions}
\label{Sec:FTF_B_Fun}

The remarkably compact explicit expressions for overlap, nuclear
attraction, and Coulomb integrals published in
\cite{Filter/Steinborn/1978b,Filter/Steinborn/1978a} undeniably
constituted a major achievement which provided considerable evidence that
$B$ functions indeed play a special role in the theory of multicenter
integrals of exponentially decaying functions.

But many open questions remained. Filter had to develop in his PhD thesis
\cite[Sections 7 - 9]{Filter/1978} some very sophisticated, but also very
complicated mathematical techniques in order to obtain the remarkably
simple expressions for the integrals mentioned above. It was by no means
clear whether and how these techniques could also be used profitably in
the case of more complicated multicenter integrals.

There was also the problem that in the Steinborn group the Fourier
transform method, which already at that time had been considered to be
one of the most or even the most important techniques for the evaluation
of multicenter integrals, had for a long time a very bad reputation and
was wrongly believed to be not very useful. This unjustified prejudice
was, however, shattered when I showed that $B$ functions possess a
Fourier transform of exceptional simplicity:
\begin{align}
  \label{FT_B_Fun}
  & \bar{B}_{n,\ell}^{m} (\alpha, \bm{p}) \; = \; (2\pi)^{-3/2} \, \int
  \, \mathrm{e}^{- \mathrm{i} \bm{p} \cdot \bm{r}} \, B_{n,\ell}^{m}
  (\alpha, \bm{r}) \, \mathrm{d}^3 \bm{r}
  \notag \\
  & \qquad \; = \; 
    (2/\pi)^{1/2} \, \frac{\alpha^{2n+\ell-1}}{[\alpha^2 +
    p^2]^{n+\ell+1}} \, \mathcal{Y}_{\ell}^{m} (- \mathrm{i} \bm{p}) \, .
\end{align}
This is the most consequential and also the most often cited result of my
PhD thesis \cite[Eq.\ (7.1-6) on p.\ 160]{Weniger/1982}. Later,
(\ref{FT_B_Fun}) was published in \cite[Eq.\
(3.7)]{Weniger/Steinborn/1983a}. Independently and almost simultaneously,
(\ref{FT_B_Fun}) was also derived by Niukkanen \cite[Eqs.\ (57) -
(58)]{Niukkanen/1984c}.

The exceptionally simple Fourier transform (\ref{FT_B_Fun}) gives $B$
functions a unique position among exponentially decaying functions. It
explains why other exponentially decaying functions like Slater-type
functions with integral principal quantum numbers, bound state hydrogen
eigenfunctions, and other functions based on generalized Laguerre
polynomials can all be expressed in terms of finite linear combinations
of $B$ functions (details and further references can be found in
\cite[Section IV]{Weniger/1985} or \cite[Section 4]{Weniger/2002}).

As remarked above, the derivation of the remarkably compact explicit
expressions for overlap, nuclear attraction, and Coulomb integrals of $B$
functions published in
\cite{Filter/Steinborn/1978b,Filter/Steinborn/1978a} was fairly
complicated and required considerable mathematical skills. If however,
the Fourier transform (\ref{FT_B_Fun}) is used, then the derivation of
these expressions is almost trivial
\cite{Weniger/Grotendorst/Steinborn/1986b}.

The Fourier transform (\ref{FT_B_Fun}) of $B$ functions is also a
convenient starting point for the evaluation of more complicated
multicenter integrals of $B$ functions by a combination of the Fourier
transform method with numerical quadratures. This was first done in
articles by Trivedi and Steinborn \cite{Trivedi/Steinborn/1983},
Grotendorst and Steinborn
\cite{Grotendorst/Steinborn/1985,Grotendorst/Steinborn/1988}, Homeier and
Steinborn
\cite{Homeier/Steinborn/1990,Homeier/Steinborn/1991,Homeier/Steinborn/1992a},
and Steinborn and Homeier \cite{Steinborn/Homeier/1990}, and in the PhD
theses of Grotendorst \cite{Grotendorst/1985} and Homeier
\cite{Homeier/1990}.

The key problem of this approach is that evaluation by numerical
quadrature can become prohibitively difficult because of the highly
oscillatory nature of the integrands. In recent years, Safouhi and
coworkers greatly enhanced the efficiency of this approach by combining
quadrature rules with suitable convergence acceleration techniques (see
for example \cite{Safouhi/2002a,Berlu/Safouhi/2003a,Berlu/Safouhi/2003b,%
  Slevinsky/Safouhi/2008} and references therein). This may well be the
currently most promising approach for the evaluation of complicated
molecular multicenter integrals of exponentially decaying functions.

\typeout{==> Section: Bessel Potential Spaces}
\section{Bessel Potential Spaces}
\label{Sec:BesselPotentialSpaces}

On the basis of our current level of understanding, we can safely claim
that the exceptionally simple Fourier transform (\ref{FT_B_Fun}) is the
most important property of $B$ functions. It fully explains the
advantageous features of $B$ functions in the context of multicenter
integrals of exponentially decaying functions. Nevertheless, it is
certainly legitimate to wonder why it took such a long time to discover
both $B$ functions and their Fourier transform (\ref{FT_B_Fun}).

For example, Geller \cite{Geller/1963a,Geller/1964a,Geller/1964b} had
studied in his work on multicenter integrals of Slater-type functions
certain types of radial integrals, which contain the radial parts of
Fourier integral representations of $B$ functions as special cases. In my
opinion, it is hard to understand that neither Geller nor anybody else
realized that these special cases deserve special attention because they
represent functions with highly useful features.

I found out later that the Fourier transform (\ref{FT_B_Fun}) of $B$
functions had at least partly been known before in a completely different
context. Aronszajn and Smith \cite{Aronszajn/Smith/1961} showed already
in 1961 in connection with their work on functional analytic properties
of solutions of certain inhomogeneous partial differential equations that
$[1+\vert \bm{\xi} \vert^{2}]^{-\alpha/2}$ with $\bm{\xi} \in
\mathbb{R}^{n}$ and $\alpha > 0$ is the $n$-dimensional Fourier transform
of what we now call a reduced Bessel function with argument $\vert \bm{x}
\vert$ with $\bm{x} \in \mathbb{R}^{n}$.

Their starting point was the following $n$-dimensional generalized
modified Helmholtz equation:
\begin{equation}
  \label{n-dimGenModHelmhotzEq}
  \bigl[1 - \nabla^{2} \bigr]^{\alpha/2} \, u (\bm{x}) 
  \; = \; g (\bm{x}) \, .
\end{equation}
Here, $u$, $g \colon \mathbb{R}^{n} \to \mathbb{C}$ are functions,
$\bm{x} \in \mathbb{R}^{n}$, $\alpha > 0$, and $\nabla^{2}$ is the
$n$-dimensional Laplacian.

Let us assume that $u$ and $g$ permit $n$-dimensional Fourier
transformation. Then,
\begin{align}
  \bar{u} (\bm{\xi}) & \; = \; (2\pi)^{-n/2} \int \, 
  \mathrm{e}^{-\mathrm{i} \bm{\xi} \cdot \bm{x}} \, u (\bm{x}) \,
  \mathrm{d}^{n} \bm{x} \, ,
  \\
  \bar{g} (\bm{\xi}) & \; = \; (2\pi)^{-n/2} \int \, 
  \mathrm{e}^{-\mathrm{i} \bm{\xi} \cdot \bm{x}} \, g (\bm{x}) \,
  \mathrm{d}^{n} \bm{x} \, ,
\end{align}
and the Fourier transform of $\bigl[1 - \nabla^{2} \bigr]^{\alpha/2}$ is
the function $[1+\vert \bm{\xi} \vert^{2}]^{\alpha/2}$. Thus, the partial
differential equation (\ref{n-dimGenModHelmhotzEq}) can be reformulated
as follows:
\begin{equation}
  \label{FourIntEq}
  \bar{u} (\bm{\xi}) \; = \; 
  \frac{\bar{g} (\bm{\xi})}{[1+\vert \bm{\xi} \vert^{2}]^{\alpha/2}} \, .  
\end{equation}
Consequently, for a given inhomogeneity $g (\bm{x})$ the unknown function
$u (\bm{x})$ can be expressed as an inverse Fourier integral:
\begin{equation}
  \label{FourIntRep}
  u (\bm{x}) \; = \; 
  (2\pi)^{-n/2} \int \, \mathrm{e}^{\mathrm{i} \bm{x} \cdot \bm{\xi}} \, 
  \frac{\bar{g} (\bm{\xi})}{[1+\vert \bm{\xi} \vert^{2}]^{\alpha/2}} \, 
  \mathrm{d}^{n} \bm{\xi} \, .  
\end{equation}
The inverse Fourier integral on the right-hand side of (\ref{FourIntRep})
can be expressed as the convolution integral,
\begin{equation}
  \label{Conv_G_g}
  u (\bm{x}) \; = \; (2\pi)^{-n/2} \int \, 
  G_{\alpha} (\bm{x}-\bm{y}) \, g (\bm{y}) \, \mathrm{d}^{n} \bm{y} \, ,
\end{equation}
where the so-called Bessel potential $G_{\alpha} (\bm{x})$ is defined as
follows:
\begin{equation}
  \label{Def:G_alpha}
  G_{\alpha} (\bm{x}) \; = \; (2\pi)^{-n/2} \int \, \frac
  {\mathrm{e}^{\mathrm{i} \bm{x} \cdot \bm{\xi}}}
  {[1+\vert \bm{\xi} \vert^{2}]^{\alpha/2}} \, 
  \mathrm{d}^{n} \bm{\xi} \, .
\end{equation}
Aronszajn and Smith \cite[Eq.\ (2, 10) on p.\ 414]{Aronszajn/Smith/1961}
showed that $G_{\alpha} (\bm{x})$ is essentially a reduced Bessel
function (compare also Samko's book \cite[pp.\ 184 - 185]{Samko/2002} and
references therein):
\begin{equation}
  \label{G_alpha_AS}
  G_{\alpha} (\bm{x}) \; = \; \frac
  {\vert \bm{x} \vert^{(\alpha-n)/2} \,
   K_{(n-\alpha)/2} (\vert \bm{x} \vert)}
  {2^{-1+\alpha/2} \, \Gamma (\alpha/2)} \, .
\end{equation}
Here, $K_{(n-\alpha)/2} (\vert \bm{x} \vert)$ is a modified Bessel
function of the second kind \cite[p.\ 66]{Magnus/Oberhettinger/Soni/1966}

As already discussed in Section \ref{Sec:Introduction}, it is my
conviction that science is now highly fragmented and communication
between researchers from different scientific (sub)disciplines does not
function particularly well. Bessel potentials provide convincing evidence
supporting my claim. For example, mathematicians working with Bessel
potentials seem to be completely unaware of the closely related Bessel
polynomials and reduced Bessel functions, and in the literature of
theoretical chemistry Bessel potentials have apparently been completely
ignored so far (an exception is my Habilitation thesis \cite[p.\
138]{Weniger/1994b}).

\typeout{==> Section: Outlook}
\section{Outlook}
\label{Sec:Outlook}

Because of their remarkable simplicity in the coordinate representation,
Slater-type functions have obvious advantages. However, their molecular
multicenter integrals are notoriously difficult. 

In contrast, $B$ function have a comparatively complicated structure.
Therefore, a more detailed look at their mathematical properties is
required to understand why $B$ functions have some highly advantageous
features in connection with molecular multicenter integrals. This does
not mean that their molecular multicenter integrals are necessarily
simple in absolute terms, but they can in most cases be evaluated
significantly more easily than the corresponding integrals of Slater-type
functions.

Thus, the history of $B$ functions, which is essentially a history of the
influx of mathematical knowledge into theoretical chemistry, is well
suited to study a question of more profound importance, namely the
interaction between mathematics and theoretical chemistry.

I think that I can claim with some justification that the historical
development of $B$ functions has not been a straight path, but at best a
not completely random walk. Moreover, the available information has not
always been utilized effectively, which obviously slowed down progress.
Thus, communication problems -- not only between mathematicians and
theoretical chemists, but also among theoretical chemists -- have more or
less been the rule rather than the exception.

Since theoretical chemistry is such a highly interdisciplinary field,
future progress will depend very much on our ability of minimizing the
detrimental effects of a breakdown of interdisciplinary and/or
intradisciplinary communication. The history of $B$ functions shows that
this may not be so easy.

The topics discussed in this article indicate that progress in
theoretical chemistry depends very much on the influx of previously
unknown mathematical ideas into theoretical chemistry. It would, however,
be wrong to assume that communication between mathematics and theoretical
chemistry would necessarily be a one way street. Problems from
theoretical chemistry in general and from the theory of $B$ functions in
particular can provide valuable inspiration for mathematical research. My
own research provides ample evidence that this is indeed possible.

During the work for my PhD thesis \cite{Weniger/1982}, series expansions
for multicenter integrals played a major role. Since it (too) often
happened that my series expansions converged slowly (see for example
\cite[Table II]{Weniger/Steinborn/1983b}), it was a natural idea to speed
up convergence with the help of (nonlinear) sequence transformations. To
the best of my knowledge, this was first done in 1967 by Petersson and
McKoy \cite{Petersson/McKoy/1967}. Unfortunately, I knew at that time
only linear series transformations as described in the classic, but now
outdated book by Knopp \cite{Knopp/1964}. Unfortunately. these linear
transformations turned out to be ineffective. I was completely ignorant
of the more powerful nonlinear transformations, which often accomplish
spectacular improvements of convergence.

My ignorance only changed when I did postdoctoral work at the Department
of Applied Mathematics of the University of Waterloo in Waterloo,
Ontario, Canada, where I -- inspired by Ji\v{r}\'{\i} \v{C}\'{\i}\v{z}ek
-- applied Pad\'{e} approximants and continued fractions for the
summation of divergent power series. After my return to Regensburg, I
tried to apply nonlinear transformations also to slowly convergent series
expansions for multicenter integrals. In some cases, remarkable
improvements of convergence were observed.

In order to understand better the power as well as the limitations of
nonlinear sequence transformations, I also worked on their theoretical
properties. As a by-product, I was able to derive several new
transformations. The majority of these transformations was published in
my long article \cite{Weniger/1989}, where also efficient algorithms for
the computation of sequence transformations as well as theoretical error
estimates and convergence properties are discussed.

Later, I applied sequence transformations successfully in such diverse
fields as the evaluation of molecular multicenter integrals of
exponentially decaying functions, the evaluation of special functions and
related objects, the summation of strongly divergent quantum mechanical
perturbation expansions, the prediction of unknown perturbation series
coefficients, and the extrapolation of quantum chemical crystal orbital
and cluster electronic structure calculations for oligomers to their
infinite chain limits of stereoregular \emph{quasi}-onedimensional
organic polymers. More information on my work both on and with sequence
transformations and exact references can by found in my recent
publications
\cite{Weniger/2004,Weniger/2007a,Weniger/2007d,Weniger/2008a}.

Thus, it is probably justified to claim that numerical mathematics and
scientific computing ultimately profited via cross-fertilization from the
convergence problems which I encountered during my PhD thesis and also
later.

Something similar can be said about the work of Safouhi. As already
mentioned in Section \ref{Sec:FTF_B_Fun}, he combines numerical
quadrature schemes with sequence transformations for the evaluation of
the highly oscillatory integral representations for multicenter
integrals. Over the years, Safouhi and his coworkers have worked hard and
studied numerous different sequences transformations in order to optimize
their computational approach. Initially, Safouhi was only interested in
the evaluation of molecular multicenter integrals, but recently he could
demonstrate that his techniques work very well also in the case of
extremely pathological oscillatory integrals that are predominantly of
mathematical interest \cite{Slevinsky/Safouhi/2008}.

Let me summarize. It is my conviction that theoretical chemists
interested in methodological questions should \emph{actively} seek the
collaboration with mathematicians. Both sides have good chances of
profiting from such a collaboration.

\begin{appendix}
\typeout{==> Appendix: Notation and Terminology}
\section{Notation and Terminology}
\label{App:NotationAndTerminology}

For the set of \emph{positive} and \emph{negative} integers, I write
$\mathbb{Z} = \{ 0, \pm 1, \pm 2, \ldots \}$, for the set of
\emph{positive} integers, I write $\mathbb{N} = \{ 1, 2, 3, \ldots \}$,
and for the set of \emph{non-negative} integers, I write $\mathbb{N}_0 =
\{ 0, 1, 2, \ldots \}$. The \emph{real} and \emph{complex} numbers are
denoted by $\mathbb{R}$ and $\mathbb{C}$, respectively. The set of
\emph{three-dimensional vectors} $\bm{r} = (x, y, z)$ with real
components $x, y, z \in \mathbb{R}$ is denoted by $\mathbb{R}^3$.

For the commonly occurring special functions of mathematical physics I
use the notation of Magnus, Oberhettinger, and Soni
\cite{Magnus/Oberhettinger/Soni/1966} unless explicitly stated
otherwise.

For the functions of angular momentum theory -- essentially (surface)
spherical harmonics $Y_{\ell}^{m} (\theta, \phi)$, regular solid
harmonics $\mathcal{Y}_{\ell}^{m} (\bm{r})$, and Gaunt coefficients
$\langle \ell_3 m_3 \vert \ell_2 m_2 \vert \ell_1 m_1 \rangle$ -- I use
the same notations and conventions as in \cite[Appendices B and
C]{Weniger/2005}.

Fourier transformation is used in its symmetrical form, i.e., a function
$f: \mathbb{R}^3 \to \mathbb{C}$ and its Fourier transform $\bar{f}$ are
connected by the integrals
\begin{align}
  \label{Def_FT}
  \bar{f} (\bm{p}) & \; = \; (2\pi)^{-3/2} \int \,
  \mathrm{e}^{-\mathrm{i} \bm{p} \cdot \bm{r}} \, f (\bm{r})
  \, \mathrm{d}^3 \bm{r} \, ,
  \\
  \label{Def_InvFT}
  f (\bm{r}) & \; = \; (2\pi)^{-3/2} \int \,
  \mathrm{e}^{\mathrm{i} \bm{r} \cdot \bm{p}} \, \bar{f} (\bm{p})
  \, \mathrm{d}^3 \bm{p} \, ,
\end{align}
\end{appendix}

%
%
%
\addcontentsline{toc}{section}{Bibliography}
{\small
\providecommand{\SortNoop}[1]{} \providecommand{\OneLetter}[1]{#1}
  \providecommand{\SwapArgs}[2]{#2#1}

}
%

\end{multicols}

\end{document}